\documentclass[aps,twocolumn,showpacs,preprintnumbers,amsmath,amssymb,superscriptaddress,floatfix,nofootinbib]{revtex4}

\usepackage{graphicx}
\usepackage{epsfig}
\usepackage{epstopdf}
\usepackage{hyperref}
\usepackage{amsmath}
\usepackage{amsfonts}
\usepackage{amssymb}
\newcommand{\Slash}[1]{\ooalign{\hfil/\hfil\crcr$#1$}}

\begin{document}

\title{The $K^- p \to f_1(1285) \Lambda$ reaction within an effective Lagrangian approach}
\date{\today}

\author{Ju-Jun Xie} \email{xiejujun@impcas.ac.cn}

\affiliation{Institute of Modern Physics, Chinese Academy of
Sciences, Lanzhou 730000, China}\affiliation{State Key Laboratory of
Theoretical Physics, Institute of Theoretical Physics, Chinese
Academy of Sciences, Beijing 100190, China} \affiliation{Research
Center for Hadron and CSR Physics, Institute of Modern Physics of
CAS and Lanzhou University, Lanzhou 730000, China}

\begin{abstract}

The production of the $f_1(1285)$ resonance in the reaction of $K^-
p \to f_1(1285) \Lambda$ is studied within an effective Lagrangian
approach. The production process is described by the $t$ channel
$K^{*+}$ meson exchange. My theoretical approach is based on the
results of chiral unitary theory where the $f_1(1285)$ resonance is
dynamically generated from the single channel $\bar{K} K^* - c.c.$
interaction. The total and differential cross sections of the $K^- p
\to f_1(1285) \Lambda$ reaction are evaluated. Within the coupling
constant of the $f_1(1285)$ to $\bar{K}K^*$ channel obtained from
the chiral unitary theory and a cut off parameter $\Lambda_c \sim
1.5$ GeV in the form factors, the experimental measurement can be
reproduced. This production process would provide further evidence
for the $\bar{K} K^* - c.c.$ nature of the $f_1(1285)$ state.

\end{abstract}

\maketitle

\section{Introduction}

The $f_1(1285)$ meson was discovered in the
1960s~\cite{d'Andlau:1965zz,Miller:1965zza,Dahl:1967pg,d'Andlau:1968zz}
as a bump in the $\pi K \bar{K}$ mass distribution in $\bar{p}p$
annihilations and in the reaction of $\pi^- p \to n \pi K \bar{K}$.
Its quantum numbers have been quite well established, thus, it is
now an axial-vector state with [$I^G(J^{PC}) = 0^+(1^{++})$] and
mass $M_{f_1} = 1281.9 \pm 0.5$ MeV and total decay width
$\Gamma_{f_1} = 24.2 \pm 1.1$ MeV~\cite{Agashe:2014kda}. Within
quark models, it is described as a $q\bar{q}$ state, while the
recent studies of Refs.~\cite{Lutz:2003fm,Roca:2005nm} have shown
that in addition to the well-established quark model picture, the
$f_1(1285)$ resonance can also be understood as a dynamically
generated state made from the single channel $\bar{K} K^* - c.c.$
interaction using the chiral unitary approach. The $f_1(1285)$
resonance cannot couple to other pseudoscalar-vector channels
because it has positive $G$ parity, and it cannot decay into two
pseudoscalar mesons (in principle $K \bar K$ in this case) for
parity and angular momentum conservation reasons. Furthermore, the
$f_1(1285)$ resonance is located below the $\bar{K} K^*$ threshold,
hence its observation in two-body decays is very difficult. Indeed,
the main decay channels of the $f_1(1285)$ resonance are $4\pi$
(branching ratio $=33\%$), $\eta \pi \pi$ ($52\%$), and $\pi \bar{K}
K$ ($9\%$).

In Ref.~\cite{genghigher}, the work of Ref.~\cite{Roca:2005nm} on
the pseudoscalar-vector interaction was extended to include the
higher order terms in the Lagrangian, and it was shown that the
effect of the higher order terms is negligible. The inclusion of the
higher-order kernel does not change the results obtained in
Ref.~\cite{Roca:2005nm} in any significant way, and thus, it lends
more confidence to the molecular picture of the $f_1(1285)$ state.
Using the dynamical picture, predictions for lattice simulations in
finite volume have been done in Ref.~\cite{gengfinite}. On the other
hand, the three-body decays of $f_1(1285) \to \eta \pi^0 \pi^0$ and
$f_1(1285) \to \pi K \bar{K}$ were studied using the picture that
the $f_1(1285)$ is dynamically generated from the single channel
$\bar{K} K^* - c.c.$ interaction in
Refs.~\cite{Aceti:2015zva,Aceti:2015pma}, where the theoretical
calculations are compatible with the experimental measurements. In
Ref.~\cite{Xie:2015lta}, the role of the $f_1(1285)$ resonance in
the decays of $J/\psi \to \phi \bar{K} K^*$ and $J/\psi \to \phi
f_1(1285)$ was investigated.

On the experimental side, the production and decay of the
$f_1(1285)$ resonance have been studied in the reaction of $K^- p
\to f_1(1285) \Lambda$ at an incident $K^-$ momentum of $4.2$
GeV~\cite{Gurtu:1978yv}. The total cross section for this reaction
is $11 \pm 3$ $\mu {\rm b}$ at $p_{K^-} = 4.2$ GeV. Later, in
Refs.~\cite{Falvard:1988fc,Jousset:1988ni}, the production of
$f_1(1285)$ resonance in the decays of $J/\psi \to \phi f_1(1285)
\to \phi 2(\pi^+ \pi^-)$ and $J/\psi \to \phi f_1(1285) \to \phi
\eta \pi^+ \pi^-$ were studied by the DM2 Collaboration~\footnote{It
was assumed that the signal observed in the $\eta \pi^+ \pi^-$
invariant mass distribution at $1297$ MeV is the $f_1(1285)$
resonance.}.

The $K \bar K^*$ channel is bound for the energy of the $f_1(1285)$
by about 100 MeV, hence this decay is not observed
experimentally~\cite{Agashe:2014kda}. However, in this work, I study
the production (rather than decay) of the $f_1(1285)$ resonance from
the $\bar{K} K^*$ interaction in the $K^- p \to f_1(1285) \Lambda$
reaction within the effective Lagrangian approach. From the
perspective that the $f_1(1285)$ is generated from the single
channel $\bar{K} K^* - c.c.$ interaction, the $K^- p \to f_1(1285)
\Lambda$ reaction is dominant by the $t$ channel $K^{*+}$ exchange,
and the interaction of $K^-$ and $K^{*+}$ generating the $f_1(1285)$
resonance. This process should be tied to the $\bar{K} K^* -c.c.$
nature of the $f_1(1285)$ state.

Before the end of this introduction, it will be helpful to mention
that, based on phenomenological Lagrangians, only the tree-level
diagram contributions are considered, in which the re-summation of
the Born amplitudes are not taken into account. However, the model
can give a reasonable description of the experimental measurement in
the considered energy region. Meanwhile, the present calculation
offers some important clues for the mechanisms of the $K^- p \to
\Lambda f_1(1285)$ reaction and makes a first effort to study the
production of the $f_1(1285)$ resonance in the $K^- p \to \Lambda
f_1(1285)$ reaction. This production process would provide further
evidence for the $\bar{K} K^* - c.c.$ nature of the $f_1(1285)$
state.

This paper is organized as follows. In Sec.~\ref{sec:formalism}, I
discuss the formalism and the main ingredients of the model. In
Sec.~\ref{sec:results}, I present my main results and, finally, a
short summary and conclusions are given in Sec.~\ref{sec:summary}.

\section{Formalism} \label{sec:formalism}

\begin{figure}[htbp]
\begin{center} \vspace{1cm}
\includegraphics[scale=1.]{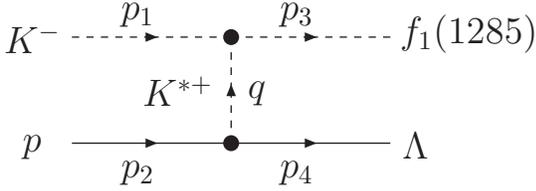}
\caption{Feynman diagram for the $K^- p \to f_1(1285) \Lambda$
reaction. It consists of $t$ channel $K^{*+}$ exchange. In this
diagram, we also show the definition of the kinematical
($p_1,~p_2,~p_3,~p_4$) variables that we used in the present
calculation. In addition, we use $q = p_3 - p_1 = p_2 - p_4$.
\label{Fig:feyn}}
\end{center}
\end{figure}

The basic tree level Feynman diagram for the $K^- p \to f_1(1285)
\Lambda$ reaction is depicted in Fig.~\ref{Fig:feyn}, where the $t$
channel $K^{*+}$ exchange is considered. In this work, the
contributions from $s$ and $u$ channels are ignored, because the
information about the $\Lambda \Lambda f_1(1285)$ and $\Lambda^*
\Lambda f_1(1285)$ vertices in the $s$ channel and $N N f_1(1285)$
and $N^*N f_1(1285)$ vertices in the $u$ channel is scarce.

To compute the contribution of the diagram shown in
Fig.~\ref{Fig:feyn}, I need the effective interaction of the
$\Lambda p K^{*+}$ vertex, which is given in terms of the
interaction Lagrangian density
by~\cite{Doring:2010ap,Ronchen:2012eg}:
\begin{eqnarray}
{\cal L}_{\Lambda N K^*} &=& - g_{\Lambda N K^*} \bar{\Lambda}
(\gamma^{\mu} -  \frac{\kappa_{K^*}}{m_{\Lambda} + m_p}\sigma^{\mu
\nu}\partial_{\nu}) K^*_{\mu} N  \nonumber \\
&& + {\rm h.c.},
\end{eqnarray}
which consists of a vector part with $\gamma^{\mu}$ and a tensor
part with $\sigma^{\mu \nu}$. Within $SU(3)$ flavor symmetry, the
value of the coupling constant, $g_{\Lambda N K^*} = -6.41$, is
obtained from the relationship $g_{\Lambda N K^*} =
-g_{NN\rho}(1+\alpha_{BBV})/\sqrt{3}$ with $g_{NN\rho} = 3.36$ and
$\alpha_{BBV} = 1.15$~\cite{Doring:2010ap,Ronchen:2012eg}. I take
$\kappa_{K^*} = 2.77$ which is obtained from $\kappa_{K^*} = 1.5
\kappa_{\rho}/(1+2\alpha_{BBV})$ with $\kappa_{\rho} =
6.1$~\cite{Machleidt:1987hj,Gasparyan:2003fp,Tsushima:2000hs,Xie:2007vs,Xie:2007qt}.

In addition to the $\Lambda N K^*$ vertex, I need also the
interaction of the $\bar{K} K^*f_1(1285)$ vertex. As mentioned
before, in the chiral unitary approach of Ref.~\cite{Roca:2005nm},
the $f_1(1285)$ resonance results as dynamically generated  from the
interaction of $\bar{K} K^* - c.c.$. One can write down the $K^-
K^{*+} f_1(1285)$ vertex of fig.~\ref{Fig:feyn} as,
\begin{eqnarray}
v = -\frac{1}{2} g_{f_1} \varepsilon^{\mu}(K^*)
\varepsilon^*_{\mu}(f_1), \label{eq:f1kbarkstar}
\end{eqnarray}
where $\varepsilon^{\mu}(K^*)$ is the polarization vector of the
$K^{*+}$ and $\varepsilon_{\mu}(f_1)$ is the polarization vector of
the $f_1(1285)$ resonance. The factor $-\frac{1}{2}$ in
Eq.~\eqref{eq:f1kbarkstar} is due to the fact that the $f_1(1285)$
resonance couples to the $I = 0$, $C = +1$ and $G = +1$ combination
of $\bar{K}$ and $K^*$ mesons, which is represented by the state
\begin{eqnarray}
\frac{1}{\sqrt{2}} (\bar{K} K^* - K \bar{K}^*) & = &
-\frac{1}{2}(K^-
K^{*+} + \bar{K}^0 K^{*0} \nonumber \\
&& - K^+ K^{*-} - K^0 \bar{K}^{*0}),
\end{eqnarray}
where I have taken $C|K^*> = - |K^*>$, which is consistent with the
standard chiral Lagrangians.

The coupling $g_{f_1}$ of the $f_1(1285)$ resonance to the $\bar{K}
K^*$ channel was obtained in Refs.~\cite{Roca:2005nm,Aceti:2015pma}
from the residue in the pole of the scattering amplitude for
$\bar{K}K^* \to \bar{K}K^*$ in $I = 0$. In the present calculation,
I take $g_{f_1} = 7555$ MeV as used in Ref.~\cite{Aceti:2015pma}.

With the effective interaction Lagrangian density for the $\Lambda N
K^*$ vertex and the $K^- K^{*+} f_1(1285)$ interaction shown in
Eq.~\eqref{eq:f1kbarkstar}, one can easily construct the invariant
scattering amplitude for the reaction of $K^- p \to f_1(1285)
\Lambda$ as,
\begin{eqnarray}
{\cal M} &=& i \frac{ g_{f_1} g_{\Lambda N K^*}}{2} \bar{u}
(p_4,s_{\Lambda}) \left [ \gamma_{\mu} +
\frac{\kappa_{K^*}}{m_{\Lambda} + m_p}(q_{\mu} - \Slash q
\gamma_{\mu}) \right ] \nonumber \\
&& u(p_2,s_p) G^{\mu \nu}_{K^*} (q) \varepsilon_{\nu}(p_3,s_{f_1})
F_1(q) F_2(q), \label{eq:amplitudeVT}
\end{eqnarray}
where $s_{\Lambda}$, $s_p$, and $s_{f_1}$ are the spin polarization
variables for the $\Lambda$, proton, and $f_1(1285)$ resonance,
respectively. The $G^{\mu \nu}_{K^*}$ is the $K^*$ meson propagator
with the form as,
\begin{eqnarray}
G^{\mu \nu}_{K^*} (q) = -i \frac{g^{\mu \nu} - q^{\mu}
q^{\nu}/m^2_{K^*}}{q^2 - m^2_{K^*}},
\end{eqnarray}
with $m_{K^*}$ the mass of the exchanged $K^{*+}$ meson. I take
$m_{K^*} = 891.66$ MeV in the present calculation.

Finally, because the hadrons are not point-like particles and the
exchanged $K^{*+}$ meson is off mass shell, one needs to include the
form factors, $F_1(q)$ and $F_2(q)$ in Eq.~\eqref{eq:amplitudeVT},
where $F_1(q)$ is for the $\Lambda N K^*$ vertex and $F_2(q)$ is for
the $K^- K^{*+} f_1(1285)$ vertex. To minimize the number of free
parameters, one can choose the same form for $F_1(q)$ and $F_2(q)$.
I adopt here the common scheme used in many previous
works~\cite{Doring:2010ap,Ronchen:2012eg,Machleidt:1987hj,Gasparyan:2003fp,Tsushima:2000hs,Xie:2007vs,Xie:2007qt}:
\begin{eqnarray}
F_1(q) = F_2(q) = \left ( \frac{\Lambda_c^2 - m^2_{K^*}}{\Lambda_c^2
- q^2} \right )^2.
\end{eqnarray}
This form has advantages and disadvantages because $(\Lambda^2_c -
m^2_{K^*})^2$ will be small and cut much if $\Lambda_c$ is not far
from the $m_{K^*}$. Actually, the value of $\Lambda_c$ can be
directly related to the hadron size. But, the question of hadron
size is still very open, one has to adjust $\Lambda_c$ to fit the
related experimental data. Empirically the cutoff parameter
$\Lambda_c$ should be at least a few hundred MeV larger than the
$K^*$ mass, hence, one can constrain it in the range of $1.3$ to
$1.7$ GeV as used in previous
works~\cite{Doring:2010ap,Ronchen:2012eg,Machleidt:1987hj,Gasparyan:2003fp,Tsushima:2000hs}
for other reactions.

In the present model, the final state interaction (FSI) is not
considered because it is difficult to treat the FSI unambiguously
due to the lack of the accurate $f_1(1285)\Lambda$ interaction. This
FSI effect would give the near threshold enhancement in the total
cross section. In this work, I do not take this FSI into account
since this estimation is rough thus it would cause uncertainty in
the model.

Then the calculation of the invariant scattering amplitude square
$|{\cal M}|^2$ and the cross section $\sigma(K^- p \to f_1(1285)
\Lambda)$ is straightforward. The differential cross section for
$K^-p \to f_1(1285) \Lambda$ at center of mass (c.m.) frame can be
expressed as
\begin{equation}
{d\sigma \over d{\rm cos}(\theta)} = \frac{1}{32\pi W^2} \frac{
|\vec{p}_3|}{|\vec{p}_1|} \left ( {1\over 2} \overline{\sum_{s_p}}
\sum_{s_{\Lambda},s_{f_1}}{|\cal M|}^2 \right ), \label{eq:dcs}
\end{equation}
where $W$ is the invariant mass of the $K^- p $ system, whereas,
$\theta$ denotes the scattering angle of the outgoing $f_1(1285)$
resonance relative to beam direction in the $\rm c.m.$ frame. In the
above equation, $\vec{p}_1$ and $\vec{p}_3$ are the 3-momenta of the
initial $K^-$ meson and the final $f_1(1285)$ mesons,
\begin{eqnarray}
|\vec{p}_1| &=& \frac{\lambda^{1/2}(W^2, m^2_{K^-},
m^2_p)}{2W}, \\
|\vec{p}_3| &=& \frac{\lambda^{1/2}(W^2, M^2_{f_1},
m^2_{\Lambda})}{2W},
\end{eqnarray}
where $\lambda(x,y,z)$ is the K\"ahlen or triangle function. The
$m_{K^-}$, $m_p$, and $m_{\Lambda}$ are the masses of the $K^-$
meson, proton, and $\Lambda$, respectively. I take $m_{K^-} =
493.68$ MeV, $m_p = 938.27$ MeV, and $m_{\Lambda} = 1115.68$ MeV.

In the effective Lagrangian approach, the sum over polarizations and
the Dirac spinors can be easily done thanks to
\begin{eqnarray}
\sum_{s_{f_1}} \varepsilon^{\mu}(p_3,s_{f_1}) \varepsilon^{\nu
*}(p_3,s_{f_1}) &=& -g^{\mu \nu} + \frac{p^{\mu}_3
p^{\nu}_3}{M^2_{f_1}}, \\
\sum_{s_p} \bar{u}(p_2,s_p) u(p_2,s_p) &=& \frac{\Slash p_2 + m_p}{2
m_p}, \\
\sum_{s_{\Lambda}} \bar{u}(p_4,s_{\Lambda}) u(p_4,s_{\Lambda}) &=&
\frac{\Slash p_4 + m_{\Lambda}}{2 m_{\Lambda}}.
\end{eqnarray}

Finally, I get
\begin{eqnarray}
&& \!\!\!\! {1\over 2} \overline{\sum_{s_p}}
\sum_{s_{\Lambda},s_{f_1}}{|\cal M|}^2 = \frac{g^2_{\Lambda N K^*}
g^2_{f_1}F^2_1(q) F^2_2(q)}{8 m_p
m_{\Lambda} (q^2 - m^2_{K^*})^2} \times \nonumber \\
&& \!\! \! \! {\rm Tr} \left ( (\Slash p_4 + m_{\Lambda})
\Gamma_{\mu} (\Slash p_2 + m_p) \Gamma_{\nu} (-g^{\mu \nu} +
\frac{p^{\mu}_3 p^{\nu}_3}{m^2_{f_1}}) \right ),
\end{eqnarray}
with
\begin{eqnarray}
\Gamma_{\mu} &=& \left ( \gamma^{\rho} +
\frac{\kappa_{K^*}}{m_{\Lambda} + m_p}(q^{\rho} - \Slash q
\gamma^{\rho}) \right ) (g_{\mu \rho} -
\frac{q_{\mu}q_{\rho}}{m^2_{K^*}}) \nonumber \\
&=& \gamma_{\mu} + \frac{\kappa_{K^*}}{m_{\Lambda} + m_p} q_{\mu} -
\frac{\kappa_{K^*}}{m_{\Lambda} + m_p} \Slash q \gamma_{\mu} -
\frac{ \Slash q q_{\mu}}{m^2_{K^*}}.
\end{eqnarray}

\section{Numerical results and discussion} \label{sec:results}

\begin{figure}[htbp]
\begin{center}
\includegraphics[scale=0.45]{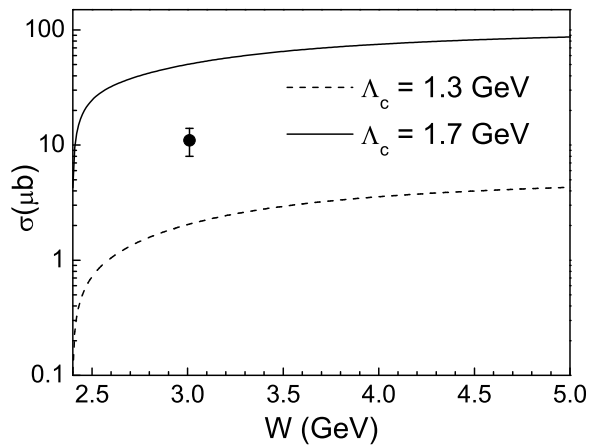}
\caption{Total cross section for the $K^- p \to f_1(1285) \Lambda$
reaction as a function of the invariant mass $W$. The experimental
data are taken from Ref.~\cite{Gurtu:1978yv}, which can be well
reproduced if I take $\Lambda_c = 1.46$ GeV.} \label{Fig:tcs}
\end{center}
\end{figure}

With the formalism and ingredients given above, the total cross
section versus the invariant mass (W)~\footnote{The $W$ can be
evaluated from the incident $K^-$ momentum, $p_{K^-}$ with the
relation $W^2 = (p_1 + p_2)^2 = m^2_{K^-} + m^2_p +
2m_p\sqrt{p^2_{K^-} + m^2_{K^-}}$.} of the $K^- p$ system for the
$K^- p \to f_1(1285) \Lambda$ reaction is evaluated. The results for
$W$ from the reaction threshold to $5.0$ GeV are shown in
Fig.~\ref{Fig:tcs} together with the experimental
data~\cite{Gurtu:1978yv} for comparison. In Fig.~\ref{Fig:tcs}, the
dashed and solid curves represent the theoretical results obtained
with $\Lambda_c = 1.3$ and $1.7$ GeV, respectively. One can see that
the experimental data can be reproduced with a reasonable value of
the cut off parameter $\Lambda_c$. Besides, the calculated cross
section is sensitive to the value of cutoff parameter used in the
form factors. The obtained results for the total cross section
$\sigma$ at $W = 3.01$ GeV are $2.1 ~\mu$b and $50.5~\mu$b with
$\Lambda_c = 1.3$ and $1.7$ GeV, respectively. To have a reliable
prediction for the cross section for the reaction $K^- p \to
f_1(1285) \Lambda$ thus requires a good knowledge of the form
factors. More and accurate experimental data can be used to
constraint the value of the cut off parameter.

On the other hand, since there is only one available experimental
datum, one will always reproduce the experimental data, $\sigma = 11
\pm 3 ~\mu$b at $W = 3.01$ GeV, by adjusting the cut off parameter
$\Lambda_c$ with a fixed coupling $g_{f_1} = 7555$ MeV. Indeed, if
one chooses $\Lambda_c = 1.46$ GeV, $\sigma = 10.9~\mu$b can be
obtained at $W = 3.01$ GeV.

\begin{figure}[htbp]
\begin{center}
\includegraphics[scale=0.45]{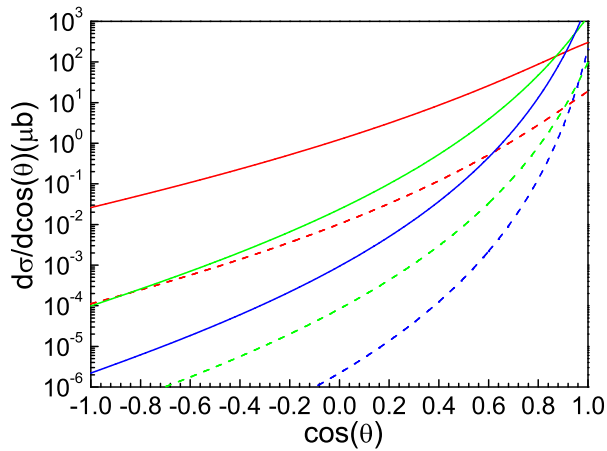}
\caption{(Color online) The $K^- p \to f_1(1285) \Lambda$
differential cross sections at different energies. The red, green,
and blue curves are obtained at $W = 3.0$, $4.0$, and $5.0$ GeV,
while the dashed and solid curves are obtained with $\Lambda_c =
1.3$ and $1.7$ GeV, respectively. \label{Fig:dcs}}
\end{center}
\end{figure}

In addition to the total cross section, I calculate also the
differential cross section for the $K^- p \to f_1(1285) \Lambda$
reaction as a function of ${\rm cos}(\theta)$ at different energies.
i.e. $W=3.0$, $4.0$, and $5.0$ GeV. The theoretical results are
shown in Fig.~\ref{Fig:dcs}. Since I have considered the
contributions from only the $t$ channel $K^{*+}$ exchange which will
give dominant contributions at the forward angle, the differential
cross sections have strong diffractive behavior, and it is stronger
when the energies are increased. These distributions can be measured
by experiment at corresponding energies. It should be pointed out
that, if there are contributions from $s$ and $u$ channels, they
will cause nondiffractive effects at off-forward angles which can be
measured directly.

In brief, either a detailed scan of the total cross section or the
angular distributions for the reaction of $K^- p \to f_1(1285)
\Lambda$ will test the model calculation, and will give more
valuable information about the mechanism of this reaction. Note that
if there were contributions from the $s$ channel terms, there will
be a clear bump (or peak) in the total cross section. Indeed, in
Ref.~\cite{Gurtu:1978yv} it was claimed that they found also some
backward contributions to the $K^- p \to f_1(1285) \Lambda$
reaction. However, the limited experimental data of
Ref.~\cite{Gurtu:1978yv} were obtained in the 1970s and only a few
signal events were observed. The future experimental observation of
the total and differential cross sections would provide very
valuable information on the reaction mechanism of $K^- p \to
f_1(1285) \Lambda$.

\section{Summary} \label{sec:summary}

In summary, the production of the $f_1(1285)$ resonance in the $K^-
p \to f_1(1285) \Lambda$ reaction is studied within an effective
Lagrangian approach. The production process is described by the $t$
channel $K^{*+}$ meson exchange, while the coupling constant of
$f_1(1285)$ to $K^- K^{*+}$ is adopted from the results of chiral
unitary theory where the $f_1(1285)$ resonance is dynamically
generated from the single channel $\bar{K} K^* - c.c.$ interaction.
The total and differential cross sections are calculated which can
be tested by future experiments. Note that the theoretical results
have a strong dependence on the cutoff parameter $\Lambda_c$ in the
form factors. To have a reliable prediction for the cross section we
need a good knowledge of the form factors.

Finally, I would like to stress that, thanks to the important role
played by the $t$ channel $K^{*+}$ exchange in the $K^- p \to
f_1(1285) \Lambda$ reaction, one can reproduce the available
experimental data with a reasonable value of the cut off parameter
in the form factors. More and accurate data for this reaction will
provide more valuable information on the reaction mechanism of $K^-
p \to f_1(1285) \Lambda$ reaction and can be used to test the model
calculation which should be tied to the $\bar{K} K^* -c.c.$ nature
of the $f_1(1285)$ state. This work constitutes a first step in this
direction.

\section*{Acknowledgments}

This work is partly supported by the National Natural Science
Foundation of China under Grant No. 11475227. It is also supported
by the Open Project Program of State Key Laboratory of Theoretical
Physics, Institute of Theoretical Physics, Chinese Academy of
Sciences, China (No.Y5KF151CJ1).

\bibliographystyle{plain}

\end{document}